\documentclass[fleqn,usenatbib]{mnras}

\usepackage[T1]{fontenc}
\usepackage{ae,aecompl}

\usepackage{physics}
\usepackage{cleveref}

\usepackage{graphicx}	
\usepackage{amsmath}	
\usepackage{amssymb}	

\usepackage[dvipsnames]{xcolor}
\usepackage[normalem]{ulem}
\usepackage{subfig}
\usepackage{tablefootnote}

\usepackage{newtxtext,newtxmath}

\graphicspath{{./figures/}}

\title[Radius valley around M-dwarfs]{Properties of the radius valley around low mass stars: Predictions from the core-powered mass-loss mechanism}

\author[A. Gupta, L. Nicholson and H.E. Schlichting]{
Akash Gupta$^{1}$\thanks{E-mail: akashgpt@ucla.edu}, Lorraine Nicholson$^{1}$ and
Hilke E. Schlichting$^{1}$
\\
$^{1}$Department of Earth, Planetary, and Space Sciences, University of California, Los Angeles, CA 90095, USA
}
\date{Accepted XXX. Received XXX; in original form XXX}

\pubyear{2022}

\begin{document}
\label{firstpage}
\pagerange{\pageref{firstpage}--\pageref{lastpage}}
\maketitle

\begin{abstract}

In recent years, analyzing the bimodality in the size distribution of small planets, i.e., the `radius valley', has given us unprecedented insight into the planet formation process. Here we explore the properties of the radius valley for low mass stars, assuming that the core-powered mass-loss is the dominant process shaping the small exoplanet population. We show that the slope of radius valley in the planet size-orbital period space, to first-order, does not vary with stellar mass and has a negative slope of $\text{d log}R_p/\text{d log}P \simeq -0.11$ even for stars as small as 0.1 $M_\odot$, as observed in latest studies. Furthermore, we find that the slope of the radius valley in the planet size-stellar mass space is $\text{d log}R_p/\text{d log}M_\ast \simeq (3 \zeta - 2)/36$ where $\zeta$ is given by the stellar mass-luminosity relation $L_\ast \propto M_\ast^\zeta$. Because $\zeta$ is $\gtrsim$ 2 and increases with stellar mass, we predict that the radius valley has a positive slope in the planet size-stellar mass space across FGKM dwarfs. This slope, however, decreases (increases) in magnitude towards lower (higher) mass stars, due to the variation of $\zeta$ with stellar mass. While around 1.0 $M_\odot$ stars the slope is $\text{d log}R_p/\text{d log}M_\ast \sim 0.37$, it is as low as $\sim 0.13$ around 0.1 $M_\odot$ stars. In addition, we find that the radius valley is narrower and less empty around lower mass stars. Finally, we show that predictions for the radius valley for core-powered mass-loss and photoevaporation become increasingly distinct for lower mass stars.

\end{abstract}




\section{Introduction}

The field of planetary sciences has witnessed a revolution over the last two decades during which over 5000 planets have been discovered. These discoveries have given us unprecedented insight into the formation and evolution of planets. Today we know that planets are quite common in our galactic neighborhood, especially small planets such as those smaller than the size of Neptune ($\sim 4 R_\oplus$). Roughly every other star hosts one such planet inside an orbital period of a 100 days \citep[e.g.,][]{borucki2010a,batalha2011a}. Furthermore, among the population of small planets, it has been noted that, surprisingly, there are very few planets of intermediate sizes $\sim$ 1.5 to 2.0 Earth radii \citep[e.g.,][]{owen2013a,fulton2017a,fulton2018a,vaneylen2018a}. In addition, planets smaller than 1.5 Earth radii (`super-Earths') have been observed to have high densities based on which, models of planetary structure infer that they should have Earth-like composition \citep[e.g.,][]{marcy2014a,rogers2015a,dressing2015a}. Whereas, planets larger than $\sim$ 1.6 Earth radii (`sub-Neptunes') have been observed to have much lower bulk densities which are consistent with large (few \% by mass) H/He atmospheres or significant fractions of ices \citep[e.g.][]{lopez2014a,jontof-hutter2016a}. This bimodality in the size and composition  distribution of small exoplanets has come to be known as the radius valley.

There are two schools of thought on the origin of the radius valley \citep[e.g.][]{bean2021a}. According to one set of theories, namely, photoevaporation and core-powered mass-loss, most of the planets we observe today formed in the presence of a protoplanetary disk and thus accreted primordial H/He envelopes \citep[][]{owen2013a,lopez2013a,jin2014a,owen2017a,ginzburg2018a,gupta2019a,gupta2020a,gupta2021a}. Some of these planets however lost their atmospheres entirely over time and became super-Earths. In contrast, planets that were able to retain their atmospheres are today's sub-Neptunes. In the case of photoevaporation, atmospheric mass-loss is driven by the high-energy radiation (e.g., XUV) from the host stars, whereas in core-powered mass-loss model, energy from the planet's formation process together with the host star's bolometric luminosity drives the atmospheric escape.

Alternatively, studies have also suggested that the radius valley could be primordial \citep[e.g.,][]{zeng2019a,mousis2020a,lee2020a}. For instance, \citet{lee2020a} suggest that super-Earths are planets that never accreted any significant atmosphere, whereas \citet{zeng2019a} suggest that sub-Neptunes are predominantly icy planets with $\sim 50\%$ ice-fraction by mass thereby explaining their larger radii. The latter theory, however, cannot yet explain radius valley features such as trends with orbital period and/or stellar mass.

In this work, we solely focus on the implications of the core-powered mass-loss theory which can explain a multitude of observations pertinent to the radius valley such as the location of the radius valley in planet size, the relative occurrences of super-Earths and sub-Neptunes, the timescales on which super-Earths form, and the slopes in the planet size-orbital period space and planet size-stellar mass space \citep[e.g.][]{fulton2018a,vaneylen2018a,martinez2019a,berger2020b}. In addition, the pertinent studies also inferred that the primordial core-mass distribution likely peaks around 4 $M_\oplus$ and that the majority of planets have cores with rocky, Earth-like composition \citet[][]{ginzburg2018a,gupta2019a}.

Previously, core-powered mass-loss studies explored the role of this mechanism in the evolution of planets around FGK dwarfs only. The motivation for these studies were the unprecedented observations from the \textit{Kepler} space observatory. A lot of the recent work in the exoplanet community has however been put into observing planets around M-dwarfs and thus also on studying the planet demographics around such stars \citep[e.g.,][]{wu2019a,cloutier2020a,vaneylen2021a,petigura2022a}. There are a multitude of reasons for this. To list a few, not only are M-dwarfs the most abundant stars in the Universe, accounting for an estimated three-quarters of the stars in our Milky Way galaxy, but they are estimated to have the highest small-planet occurrence rate among FGKM dwarfs for a given range of orbital periods \citep[e.g.][]{hsu2020a}. Furthermore, it is easier to observe small planets around these smaller stars, and many of them could in fact be in the habitable zones as these stars are much cooler \citep[e.g.][]{scalo2007a}. The TRAPPIST-1 system is an ideal example of a planetary system around an ultra-low mass star (0.089 M$_\odot$) that hosts several planets in the habitable zone \citep[][]{gillon2016a,gillon2017a}. Furthermore, as alluded to in \citet{rogers2021a}, looking at a wider range of stellar masses will facilitate our efforts towards distinguishing between the expected signatures of photoevaporation and core-powered mass-loss imprinted on the exoplanet population. Given these reasons, in this paper, we investigate the evolution of sub-Neptunes and super-Earths around low-mass stars with masses as low as 0.08 $M_\odot$.

Recent observational studies on planet demographics around such low mass stars show that the radius valley extends at least to early-M dwarfs \citep[e.g.,][]{cloutier2020a,vaneylen2021a,petigura2022a}. To characterize the radius valley, studies have tried to estimate its location in planet size and slope in different parameter spaces such as planet size-orbital period and planet size-stellar mass. \citet{vaneylen2021a} and \citet{petigura2022a} recently analyzed the distribution of planets around M and K dwarfs with masses $\sim$ 0.15-0.6 $M_\odot$ and 0.5-0.7 $M_\odot$, respectively. They found that the radius valley in the planet size-orbital period space has a negative slope of -0.11$^{+0.05}_{-0.04}$ and -0.12$^{+0.02}_{-0.04}$, respectively. These values are consistent with those reported previously for FGK dwarfs $\sim$ 0.8-1.2 $M_\odot$ \citep[e.g.][]{fulton2018a,vaneylen2018a,martinez2019a}. Theories based on atmospheric escape show that they can explain this slope. For instance, using simple analytical arguments and numerical simulations, \citet{gupta2019a} showed that if atmospheric loss is dominated by core-powered mass-loss, the radius valley slope for planets around FGK stars in the planet size-orbital period space should be $-0.11$. Contrary to the observational studies mentioned above, \citet{cloutier2020a} measured a positive slope of 0.058$\pm$0.022 for the radius valley in the same parameter space for M dwarfs ranging 0.08-0.6 $M_\odot$. The authors attributed this to an increasing role of gas-poor accretion around lower mass stars \citep{lee2016a,lopez2018a}. 

In this study, we simulate planets around these lower stellar masses and show that if planet evolution is indeed dictated by core-powered mass-loss, the slope of the radius valley in the planet size-orbital period space should remain negative and close to -0.11 as predicted previously for Sun-like stars in \citet{gupta2019a}.

Furthermore, while it has been already known that the radius valley around dwarfs has a positive slope in the planet size-stellar mass space \citep[][]{fulton2018a}, recent observational studies on planets around early M to FGK dwarfs have tried to quantify this slope. \citet{petigura2022a} looked at 0.5-1.4 $M_\odot$ stars and estimated a slope of 0.18$^{+0.08}_{-0.07}$ whereas \citet{berger2020b} found a slope of 0.26$^{+0.21}_{-0.16}$ for a similar stellar mass range but a distinct dataset. These studies have then compared these measurements with theoretical predictions from photoevaporation and core-powered mass-loss models and have noted how with the current precision, both theories are consistent within 1-2 $\sigma$. There are however caveats with making such comparisons. For instance, the predicted radius valley slope in the planet size-stellar mass space of 0.33 reported in \citet{gupta2020a} was for a specific subset of the \textit{California-Kepler Survey} (CKS) dataset then available that consisted of FGK dwarfs only. This is important because this slope depends on how strongly or weakly the stellar mass and luminosity are correlated due to the dependence of mass-loss rate on planetary equilibrium temperature and thus stellar luminosity, and this correlation itself depends on stellar mass \citep[e.g.,][]{salaris2005a,eker2018a,berger2020b}. For a unique dataset of stars, the estimated slope will thus be unique. In this work, we show how this correlation between the stellar mass and luminosity changes with stellar mass, and ultimately, present predictions for how the radius valley changes towards lower stellar masses.

This paper is structured as follows: In \Cref{sec:Methods}, we review how we model the structure of a planet and its evolution over time under the core-powered mass-loss mechanism. Following this, we describe the changes in our planet evolution model in comparison to previous studies and how we model the exoplanet population. We then discuss our results in \Cref{sec:Results} and show how the radius valley around M dwarfs differs from that around FGK dwarfs. Finally, we present our conclusions in \Cref{sec:Conclusions}.

\section{Methods} \label{sec:Methods}

In this section, we provide a brief overview of how a planet evolves after the dispersal of the protoplanetary gas disk, and the accompanying atmospheric mass-loss via the `boil-off' or `spontaneous mass-loss' process \citep[e.g.][]{owen2016a,ginzburg2016a}. While multiple theories have been put forth in this regard, we assume that any subsequent evolution is solely driven by the thermal evolution and atmospheric escape driven by the core-powered mass-loss mechanism as previously explored in \citet{ginzburg2018a,gupta2019a,gupta2020a}. 

During the formation of a planet, as solids accrete to form a planetary core, the gravitational energy resulting from this process gets converted into thermal energy. If this happens after the protoplanetary gas disk has dispersed, this energy can get radiated away on extremely short timescales. However, if this occurs during the presence of a gas disk, the planet will, once it has cooled to temperatures sufficiently low such that gas can become gravitationally bound, also accrete an H/He envelope. Once this planetary envelope becomes optically thick, any energy exchange between the surroundings and the planet will be regulated by the radiative diffusion across the radiative-convective boundary leading to much longer cooling timescales than if the planet had an optically thin or no atmosphere \citep[e.g.][]{rafikov2006a,lee2015a,ginzburg2016a}. Effectively, such an atmosphere will act as a thermal blanket for the planet allowing a significant fraction of the formation energy to be trapped within. As the protoplanetary gas disk then disperses, the decrease in the pressure support outside a planet's atmosphere will lead to atmospheric escape. This mass-loss phase is known as the boil-off or the spontaneous mass-loss process and sets the initial conditions for both core-powered mass-loss and photoevaporation. The idea behind core-powered mass-loss is that the remnant primordial formation energy can eventually unbind the planet's atmosphere via a Parker-type hydrodynamic wind in the outer regions of the atmosphere where the temperature is set by the host star's bolometric luminosity \citep[e.g.][]{parker1958a}. Planets that eventually lose their entire atmospheres are today's super-Earths whereas those that are able to retain some of their primordial H/He atmospheres are the sub-Neptunes we see today. This results in a bimodal planet size distribution and thus, a radius valley.

\subsection{Planet Structure and Evolution}\label{sec:planet_structure}

For our planet evolution model, we use the same framework as described in \citet[][]{ginzburg2018a} and \citet[][]{gupta2019a} unless specified otherwise. We assume that post-disk dispersal and boil-off/spontaneous mass-loss, a typical planet with mass $M_p$ and radius $R_p$, has a core with an Earth-like composition that is surrounded by an H$_2$ atmosphere of mass $M_{atm}$. The core here refers to the non-gaseous part of the planet that has a mass $M_c$ and radius $R_c$ and the core is assumed to account for most of the mass of the planet such that $M_c$ $\sim$ $M_p$. For the atmosphere, we model it to have an inner convective and an outer radiative region \citep[e.g.][]{piso2014a,lee2015a}. It is thus assumed that the inner region has an adiabatic profile whereas the radiative region is nearly isothermal with a temperature equal to the planetary equilibrium temperature $T_{eq}$. The radius at which such an atmosphere transitions from the convective to the radiative region is the radiative-convective boundary $R_{rcb}$. The atmosphere's density decreases exponentially with radius in the radiative region, and thus most of the mass of the atmosphere, $M_{atm}$, resides within the convective region. Consequently, the photospheric radius of the planet $R_p$ is approximately equal to $R_{rcb}$ which, in turn, is assumed to be roughly twice the core radius after the disk dispersal, based on previous work \citep[][]{ginzburg2016a}.

Given this structural model for a planet, we can now discuss how its atmospheric mass fraction $f$ = $M_{atm}/M_c$ and cooling energy $E_{cool}$, i.e. the energy that a planet can lose, evolve over time. We assume that given some initial $f$, a planet loses this atmosphere in either an energy-limited regime or a Bondi-limited regime. The energy limited regime corresponds to the absolute upper limit on atmospheric escape where all the luminosity from a cooling planet goes into driving mass loss. This mass-loss rate $\dot{M}_{atm}^E$ can be written as
\begin{equation}\label{eq:M_loss_rate_E}
    \dot{M}_{atm}^E \simeq \frac{L_{rcb}}{g R_c},
\end{equation}
where $L_{rcb}$ is the luminosity of the planet at the radiative-convective boundary and $g$ is the acceleration due to gravity at $R_{rcb}$. {In reality, the efficiency with which a planet's cooling luminosity can drive mass should be $<$100\%, but note that the contribution from the host star’s bolometric luminosity, that we don’t model in detail here, also plays a supporting role in driving this mass-loss, as it contributes to maintaining the radiative-convective profile of the planet. Ideally, detailed radiative-hydrodynamic calculations are needed to solve for atmospheric escape with incoming and outgoing radiation due to the host star and planet. This is unfortunately beyond the scope of this work. Nevertheless, as discussed previously in \citet{gupta2019a} and in this manuscript, the shape of the radius valley, i.e., its slope and depth, is primarily dictated by the balance of the planetary cooling and mass-loss timescales. Thus, we do not expect the assumption of a maximum mass-loss efficiency to have a significant implication for the slope of the radius valley and note that there is also likely a significant degeneracy between the mass-loss efficiency and a planet’s assumed initial atmosphere mass-fraction, which remains highly uncertain.}

Alternatively, the Bondi-limited mass-loss rate $\dot{M}_{atm}^B$ is the maximum mass-loss rate physically possible given the finite thermal velocities of the gas molecules at the Bondi radius which, in turn, is dictated by the bolometric luminosity of the host star and the planet's semi-major axis. In this regime, the mass-loss rate $\dot{M}_{atm}^B$, can be written as 
\begin{equation}\label{eq:M_loss_rate_B}
\dot{M}_{atm}^B = 4\pi R_s^2 c_s \rho_{s} = 4\pi R_s^2 c_s \rho_{rcb} \; \text{exp}\left( -\frac{GM
_p}{c_s^2 R_{rcb}}\right),
\end{equation}
where $R_s = G M_c/2c_s^2$ is the sonic radius, i.e., the radius at which the escaping atmosphere reaches sonic velocities, $c_s$ = $(k_B T_{eq}/\mu)^{1/2}$ is the isothermal speed of sound with Boltzmann constant $k_B$ and mean molecular weight $\mu$, and $\rho_s$ and $\rho_{rcb}$ are the atmospheric densities at $R_s$ and $R_{rcb}$. The exponential dependence of the mass-loss rate on $M_p$, $R_{rcb}$ and $T_{eq}$ stems from the isothermal nature of the outer regions of the atmosphere which requires an exponential decline in the atmosphere's density with radius. Ultimately, the mass-loss rate a planet experiences at a particular time in its evolution is the smaller of the Bondi- and energy-limited mass-loss rates, i.e., $\dot{M}_{atm}=min\{\dot{M}_{atm}^E,\dot{M}_{atm}^B\}$. We can thus express the mass-loss timescale, $t_{loss}$, of a planet as
\begin{equation}\label{eq:t_loss}
t_{loss}=\frac{M_{atm}}{|\text{d}M_{atm}/\text{d}t|} = \frac{1}{\text{min}\left\{ {\dot{M}_{atm}^E}/{M_{atm}},\; {\dot{M}_{atm}^B}/{M_{atm}} \right\}}.
\end{equation}
On the other hand, the cooling timescale for a planet can be written as
\begin{equation}\label{eq:t_cool}
t_{cool}= \frac{E_{cool}}{|\text{d}E_{cool}/\text{d}t|}=\frac{E_{cool}}{L_{rcb}}.
\end{equation}
Together, Equations \ref{eq:t_loss} and \ref{eq:t_cool} determine how a planet evolves in our model. We thus numerically solve these equations over time to ultimately track how a planet's size and atmosphere mass-fraction evolve.

\begin{figure}
\centering
    \includegraphics[width=8cm,trim=165 545 1040 235,clip]{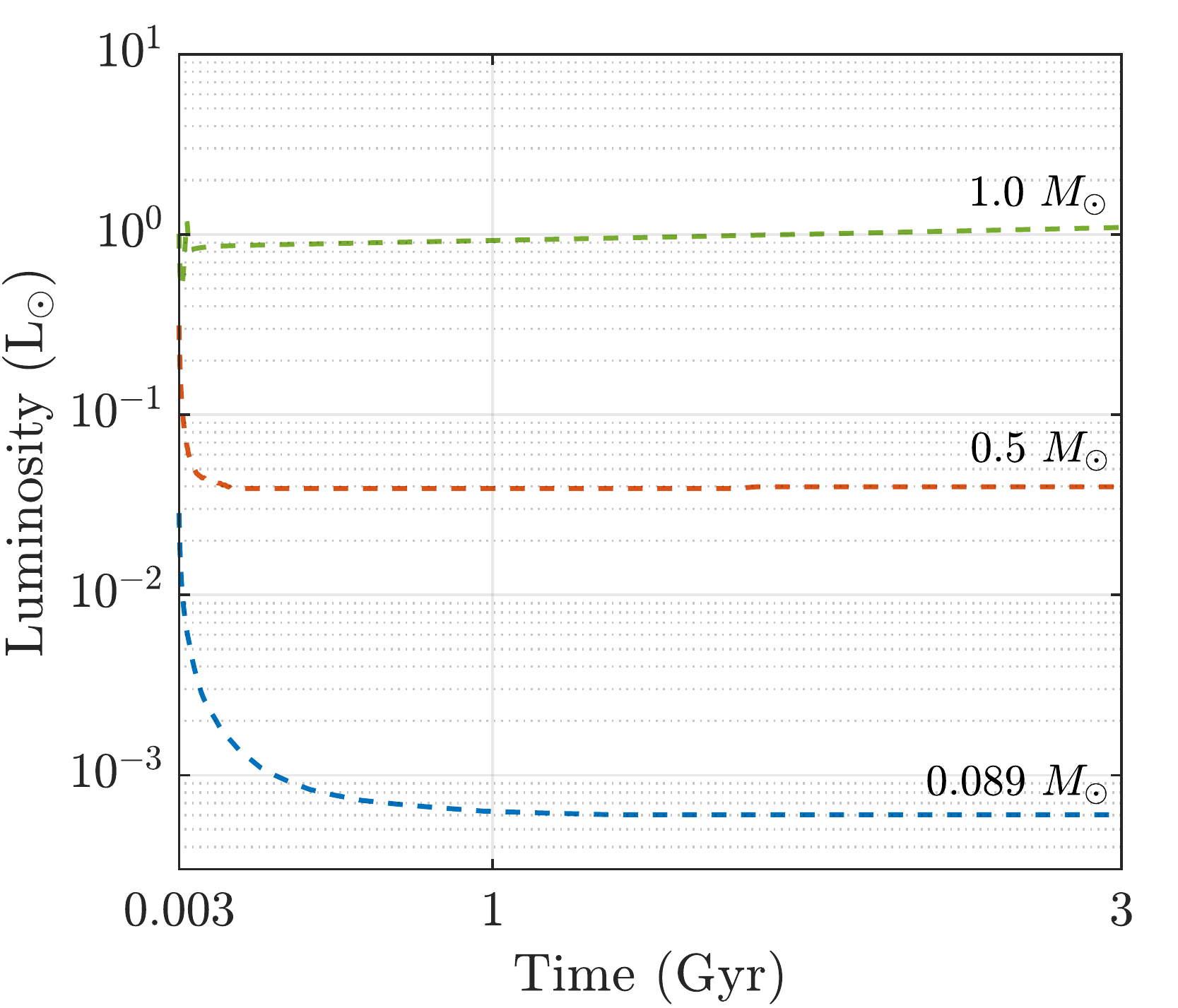}
    \caption{Evolution in the luminosities of stars of different masses from an age of 3 Myr to 3 Gyr. The plot shows that while there is an no significant change in the luminosity of a Sun-like 1.0 $M_\odot$ star, the luminosity of a star such as TRAPPIST-1 with a mass of 0.089 $M_\odot$ undergoes a change in luminosity of one-to-two orders in magnitude during the first Gyr of its lifetime.}
     \label{fig:luminosity_evolution}
\end{figure}

\subsection{Accounting for evolution of a host star's luminosity}\label{sec:lum_evolution}

Previously in \citet{ginzburg2018a} and \citet{gupta2019a,gupta2020a}, we did not account for the evolution in the luminosity of the host stars, but this change can be significant for low mass stars. This is apparent in \Cref{fig:luminosity_evolution} where we show the luminosity evolution tracks for a range of stellar masses. This figure begins at 3 Myr of age, assuming that to be the approximate time of disk dispersal \citep[e.g.][]{mamajek2009a}. We see that for a Sun-like 1 M$_\odot$ star, the luminosity undergoes a negligible change after the dispersal of its protoplanetary disk. We do note that the luminosity of such a star increases continually over the time period shown in \Cref{fig:luminosity_evolution}, but as we will see later in \Cref{sec:Results}, this has no significant impact on the core-powered atmospheric mass-loss results. On the other end of the spectrum, for a TRAPPIST-1 like ultra-low mass star, the luminosity changes by one to two orders-of-magnitude during the first Gyr of evolution. The stellar luminosity $L_\ast$ plays a key role in a planet's evolution as it sets its equilibrium temperature
\begin{equation}
T_{eq} = \left(\frac{1}{16\pi\sigma}\frac{L_*}{a^2}\right)^\frac{1}{4},
\end{equation}
where $\sigma$ is the Stefan-Boltzmann constant and $a$ is the planet's semi-major axis. As mentioned in the preceding paragraph, the mass-loss rate of a planet has an exponential dependence on $T_{eq}$. As a consequence, the mass-loss rate or mass-loss timescale, $t_{loss}$, is sensitive to changes in $T_{eq}$.

Therefore, to track planet evolution around low-mass stars, we explicitly include the evolution of the stellar luminosity with time in our core-powered mass-loss model as follows: For stars more massive than 0.1 $M_\odot$, we use the MIST stellar evolutionary tracks \citep[version 1.2;][]{dotter2016a,choi2016a} that are computed with the Modules for Experiments in Stellar Astrophysics (MESA) code \citep[][]{paxton2011a,paxton2013a,paxton2015a,paxton2018a,paxton2019a}. For stars less massive than 0.1 $M_\odot$, we directly use the MESA code (version: mesa-r21.12.1) to compute stellar evolution tracks as data for such ultra-low mass stars are not available as part of the MIST database. In addition, we assume all the stars have a metallicity of $Z=0.014$.

\begin{figure*}
\centering

\includegraphics[width=0.45\textwidth,trim=810 400 220 300,clip]{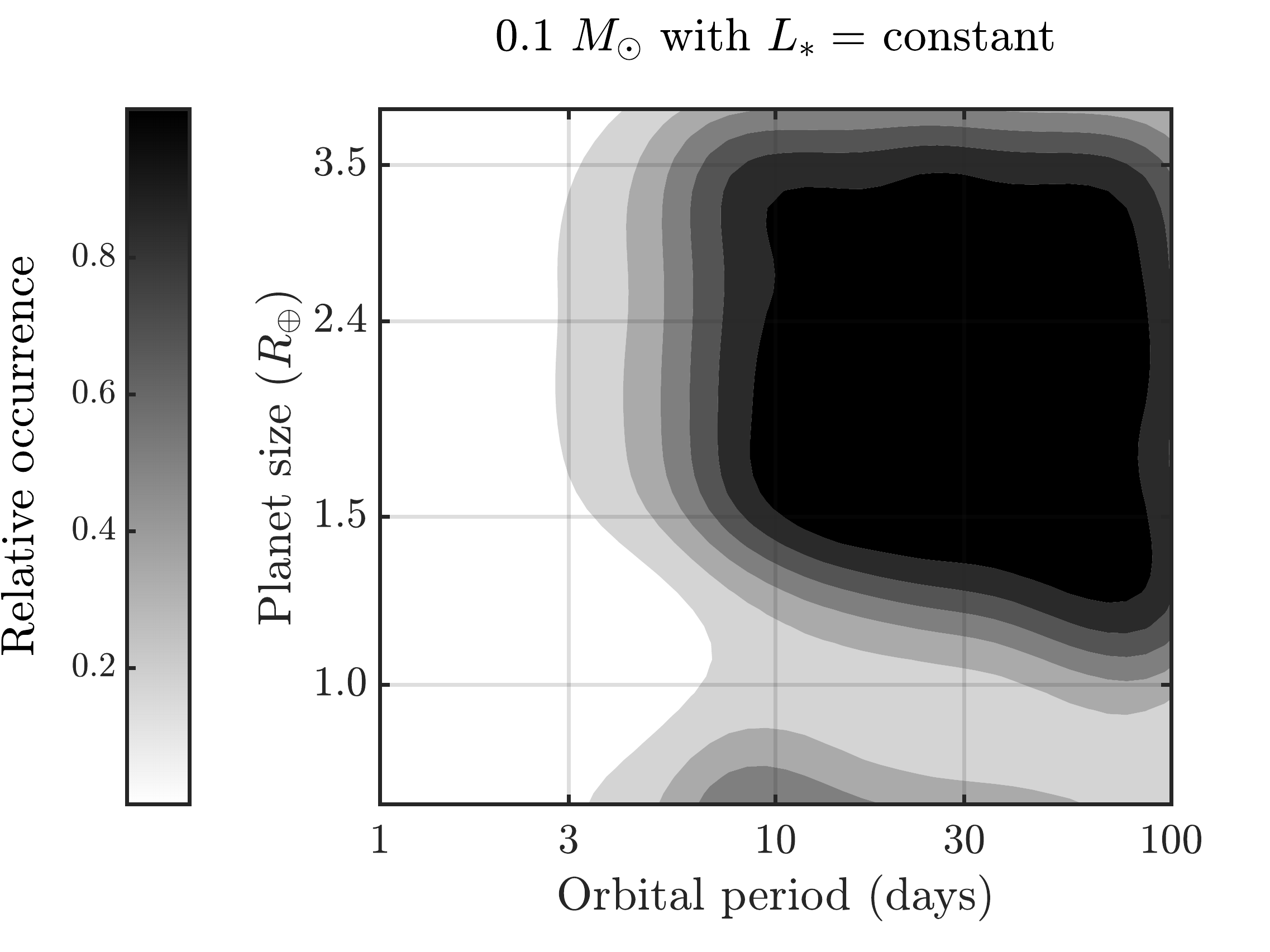}
\includegraphics[width=0.45\textwidth,trim=810 400 220 300,clip]{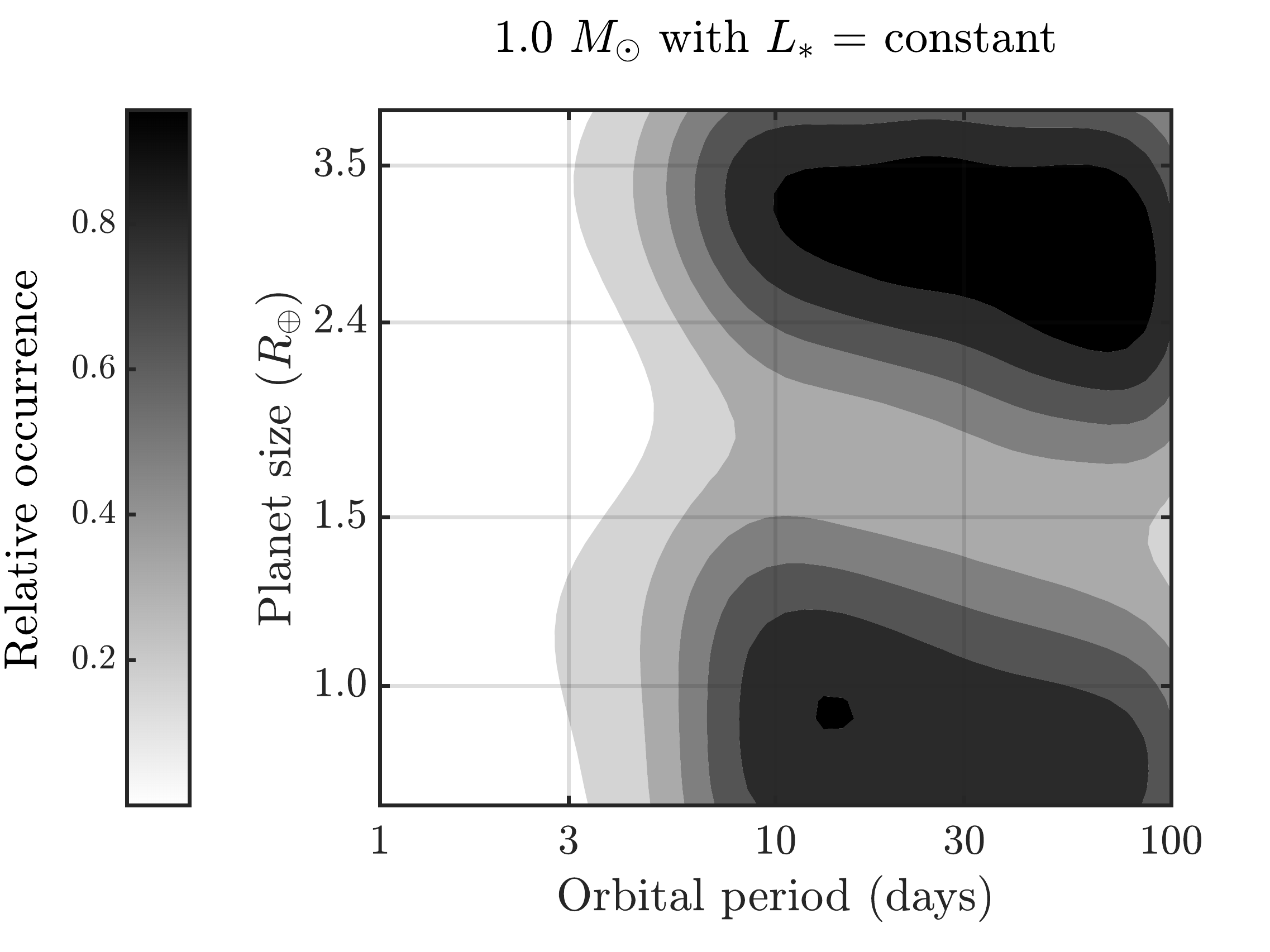}

\includegraphics[width=0.45\textwidth,trim=150 400 880 280,clip]{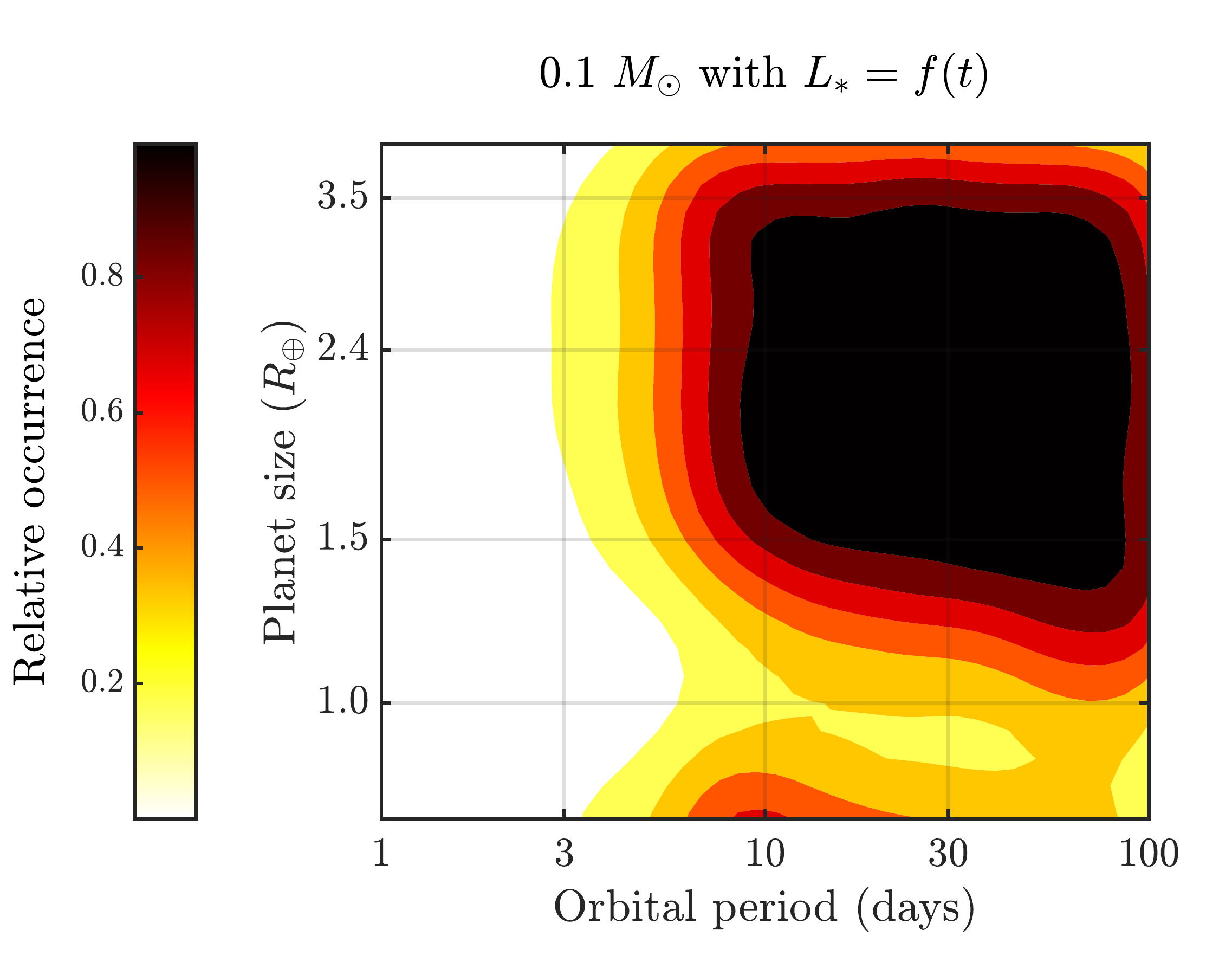} \includegraphics[width=0.45\textwidth,trim=150 400 880 280,clip]{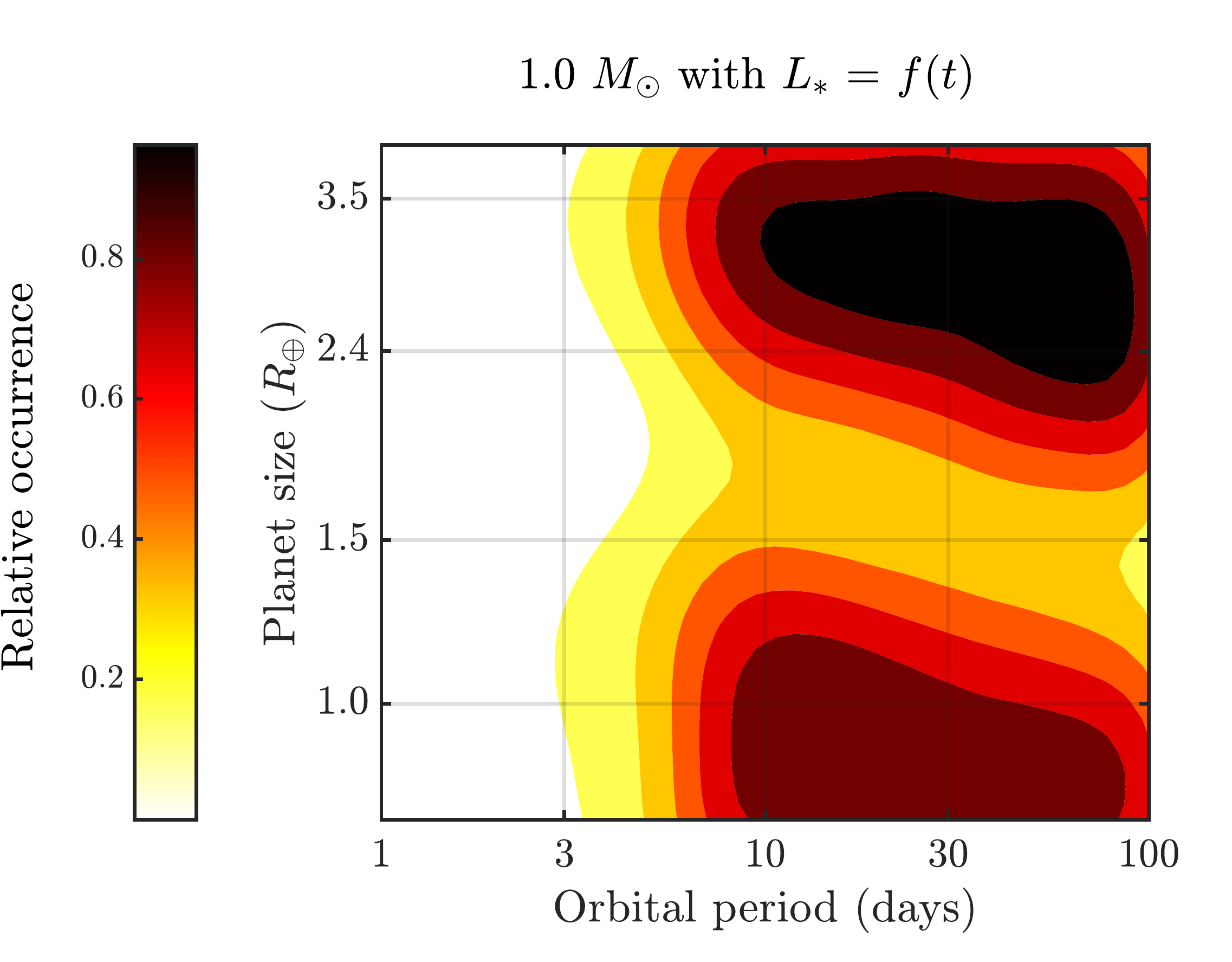}

\caption{Distribution of planets in the planet size-orbital period space. The left and right panels show planet populations around a 0.1 and 1.0 M$_\odot$ star, respectively. While the gray plots in the top row represent the case where planets were evolved around stars of constant luminosity (taken to be the luminosity at 3 Gyr), the bottom row shows the final planet size distributions when the luminosity evolution of the host stars is accounted for. Comparing the top and bottom rows shows that, unsurprisingly, the impact of the host star's luminosity evolution is negligible for 1.0 $M_\odot$ stars. On the other hand, for 0.1 $M_\odot$ stars the host star's luminosity evolution results in a narrower and less empty radius valley but, at the same time, does not lead to a change in its location or slope. These plots also show that the radius valley slope is the same around both 0.1 and 1.0 $M_\odot$ stars: $\text{d log}R_p/\text{d log}P =$ -0.11, which is consistent with observational results reported by \citet[][]{vaneylen2021a} and \citet[][]{petigura2022a}.}
 \label{fig:luminosity_and_mass_effect}
\end{figure*}

\subsection{Modeling planet population}\label{sec:ini_condns}

In this subsection, we describe how we model the distribution of atmosphere mass fractions $f$, core masses $M_c$ and orbital periods $P$. 

Following previous studies \citep[e.g.,][]{ginzburg2018a,gupta2019a}, we assume that planets have an initial atmospheric mass-fraction of 
\begin{equation}\label{eq:f}
    f = \frac{M_{atm}}{M_c} \simeq 0.05 (M_c/M_\oplus)^{1/2}.
\end{equation}
This is motivated by past studies of gas accretion and the atmosphere loss that follows the disk dispersal, i.e. the boil-off or spontaneous mass-loss process \citep[e.g.][]{owen2016a,ginzburg2016a}. {We do note however that such a one-to-one relation between $f$ and $M_c$ is not expected in nature due to the stochasticity inherent in the process of planet formation even if the correlation is physically motivated. To check if this approximation has an impact on the radius valley, we tested our evolution models with a wider distribution of atmospheric mass-fractions. Specifically, we implemented a log-normal distribution with a mean given by \Cref{eq:f} and a standard deviation of 0.25, which encompasses about a factor of 2 variations in $f$. We find that the results with the log-normal distribution are essentially indistinguishable from the ones presented in this paper. Furthermore, in \citet{gupta2019a} we demonstrated that even for a log-uniform distribution of initial-envelope mass-factions, as assumed in \citet{owen2017a}, the radius valley is still a robust outcome although the exact shape, especially of the sub-Neptune population, is slightly different.}

The period distribution has been observed to be fairly constant across a significant range of stellar masses, 0.5 M$_\odot$ to 1.4 M$_\odot$ \citep[see][]{petigura2022a}. In this work, we assume that the same observed period distribution
\begin{equation}{\label{eq:P_distr}}
    \dv{N}{\;\text{log}P} \propto \begin{cases}
         P^{2}, & {P < 8\; \text{days}} \\
        \text{constant}, & {P > 8\; \text{days}}, \text{ and} 
    \end{cases}  
\end{equation}
applies to the entire range of stellar masses explored here.

Finally, we assume that the planetary masses are distributed uniformly in log-space. This allows us to share predictions in this paper that are agnostic to the yet unknown `true' planet mass distribution and how it depends on stellar properties such as stellar mass; see \citet[][]{teske2020a}.

\section{Results \& Discussion} \label{sec:Results}

In this section, we explore the changes in the planet demographics around lower mass stars. This includes investigating the location and slope of the radius valley in planet size-orbital period and planet size-stellar mass space. Finally, we discuss how focusing on lower stellar masses can make it easier to distinguish between the signatures of core-powered mass-loss and other mechanisms that have been put forth to explain the radius valley.

\begin{figure*}
\centering
\includegraphics[width=0.80\textwidth,trim= 155 400 500 340,clip]{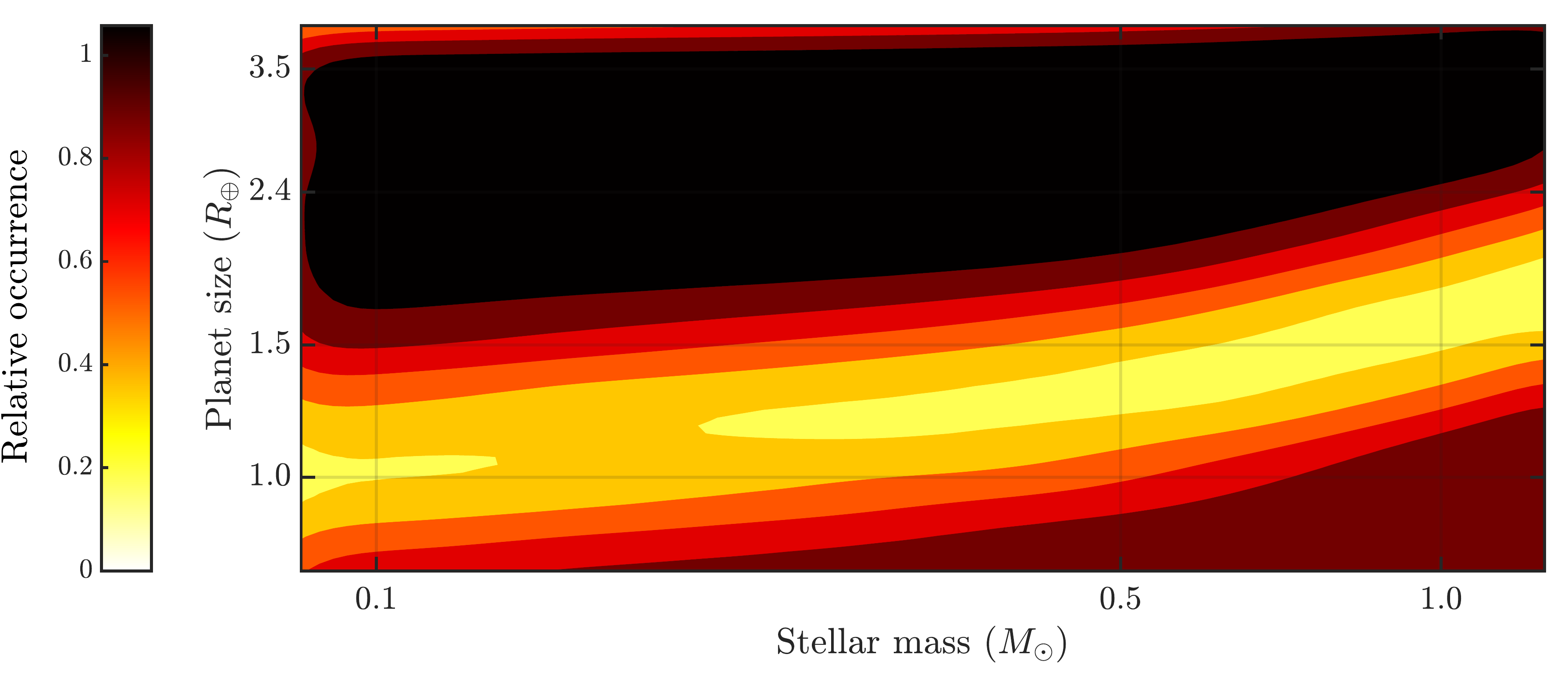}
\caption{Distribution of planets in the planet size-stellar mass space for stellar masses ranging over 0.08 to 1.25 $M_\odot$. Plot shows how the radius valley decreases in planet size with decreasing stellar masses and becomes relatively narrower and less empty. We also find that the slope of the radius valley $\text{d log}R_p/\text{d log}M_\ast$ is not constant and becomes shallower towards lower mass stars. As discussed in the main text, the radius valley slope is positive because at a given orbital period, planets around more massive stars have higher equilibrium temperatures as more massive stars are more luminuous. Therefore, a planet around a relatively massive star is more susceptible to complete atmospheric loss and the valley moves up in planet size with increasing stellar mass. This slope however changes with stellar mass because the degree of correlation between stellar mass and stellar luminosity, encapsulated in $\zeta$, itself changes with stellar mass. This change in slope is even more apparent in \Cref{fig:Rp_vs_Ms_distribution_different_ranges}.}
\label{fig:Rp_vs_Ms_distribution}
\end{figure*}

\subsection{Radius valley as a function of orbital period around low mass stars}
As discussed in \Cref{sec:lum_evolution}, in past core-powered mass-loss studies the evolution of the host star's luminosity was not included, but this change can be significant for low mass stars; see \Cref{fig:luminosity_evolution}. In \Cref{fig:luminosity_and_mass_effect}, we show the impact of this luminosity evolution for a Sun-like 1.0 $M_\odot$ star and an ultra-cool M dwarf with a mass of 0.1 $M_\odot$. This figure shows the planet size distribution in the planet size-orbital period ($R_p-P$) space after 3 Gyr of evolution given our initial conditions. For plots in the bottom row, we accounted for evolution in the stellar luminosities whereas for the top row we did not and evolved the planets at a fixed luminosity - the star's luminosity at the age of 3 Gyr. Comparing the top and bottom rows for the 1.0 M$_\odot$ case shows that the two cases are practically identical - reinforcing that it is reasonable to ignore the luminosity evolution for FGK stars in this context, as assumed in previous core-powered mass-loss studies \citep[e.g.][]{gupta2020a}. Comparing the top and bottom rows for the 0.1 M$_\odot$ case shows however that for such low mass stars the radius valley is narrower and less empty when the luminosity evolution is included. Nevertheless, the planet distributions are not drastically different, and we can thus exploit this to make analytical estimates for the slopes of radius valley in different parameter spaces down to very low stellar masses.

It is evident from \Cref{fig:luminosity_and_mass_effect} that core-powered mass-loss predicts a negative slope for the radius valley in the $R_p-P$ parameter space even around ultra-low mass stars. As previously derived in \citet{gupta2019a,gupta2020a}, the slope of the radius valley can be estimated by realizing that sub-Neptunes are planets that reached a point in their evolution where they cool and shrink faster than they lose mass while they still have substantial envelopes. Whereas super-Earths were able to lose most or all of their envelopes before cooling could catch up with the mass-loss. Therefore, the slope of the radius valley can be estimated by equating the mass loss and cooling timescales, which yields
\begin{equation}\label{eq:slope_exponential_term}
\frac{GM_p}{c_s^2 R_{rcb}} \simeq \text{constant}.
\end{equation}
as shown in \citet[][]{gupta2019a,gupta2020a}. Substituting for the speed of sound, the mass-radius relation for the core $M_c \propto R_c^4 \rho_{c\ast}^{4/3}$ where $\rho_{c\ast}$ is the density of the core scaled to an Earth mass, and using the fact that $R_{rcb} = R_p \sim 2 R_c$, simplifies \Cref{eq:slope_exponential_term} to
\begin{equation}\label{eq:slope_Rp_P}
R_p^{3} T_{eq}^{-1} \rho_{c\ast}^{4/3} \sim \text{constant}.
\end{equation}
By definition, planetary equilibrium temperature depends on stellar luminosity and semi-major axis such that $T_{eq}^{4} \propto L_\ast/a^2$. Using the law of gravitation, we know that a planet's orbital period $P^2 \propto a^3/M_\ast$. Putting these two relations together gives that $T_{eq} \propto P^{-1/3} L_\ast^{1/4} M_\ast^{-1/6}$. In addition, it is also known that the stellar mass and luminosity are positively correlated, i.e. $L_\ast \propto M_\ast^\zeta$, where $\zeta$ \citep[e.g.][]{salaris2005a,eker2018a,berger2020b}. We thus find that equilibrium temperature, orbital period and stellar mass are correlated such that $T_{eq}^{-1} \propto P^{1/3} M_\ast^{-(\zeta/4) + (1/6)}$. Substituting this in \Cref{eq:slope_Rp_P} gives us
\begin{equation}\label{eq:slope_Rp_misc}
    R_p^{3} P^{1/3} M_\ast^{-(\zeta/4) + (1/6)} \rho_{c\ast}^{4/3} \sim \text{constant}.
\end{equation}
Assuming that planets across orbital periods of 1-100 days have, on average, similar core composition, the slope in the $R_p-P$ space around a star of a particular mass is simply $\text{d log}R_p/\text{d log}P = -1/9 \simeq$ -0.11, i.e. independent of the mass of the host star. In other words, the susceptibility to atmospheric loss strongly depends on the planetary equilibrium temperature $T_{eq}$. Around any star, $T_{eq}$ decreases with increasing orbital period such that $T_{eq} \propto P^{-1/3}$. A planet at longer orbital periods is thus less susceptible to significant atmospheric loss and the location of the radius valley moves to smaller planet sizes with increasing orbital period, i.e. it has a negative slope in the $R_p-P$ space.

This result is consistent with the observations of slope across FGKM dwarfs. Recently, \citet{vaneylen2021a} and \citet{petigura2022a} reported negative slopes for the radius valley around low-mass stars: $\text{d log}R_p/\text{d log}P =$ -0.11$^{+0.05}_{-0.04}$ for stars in the range $\sim$ 0.15-0.6 $M_\odot$ and -0.12$^{+0.02}_{-0.04}$, for stars in the range 0.5-0.8 $M_\odot$, respectively. Previously, studies such as \citet{vaneylen2018a} and \citet{martinez2019a} have reported negative slopes of $\text{d log}R_p/\text{d log}P =$ $-0.09_{-0.02}^{+0.04}$ and $-0.11_{-0.03}^{+0.03}$, respectively, for the radius valley around FGK dwarfs.

\subsection{Radius valley as a function of stellar mass around low mass stars}
\begin{figure*}
\centering
\includegraphics[width=0.99\textwidth,trim= 150 400 200 340,clip]{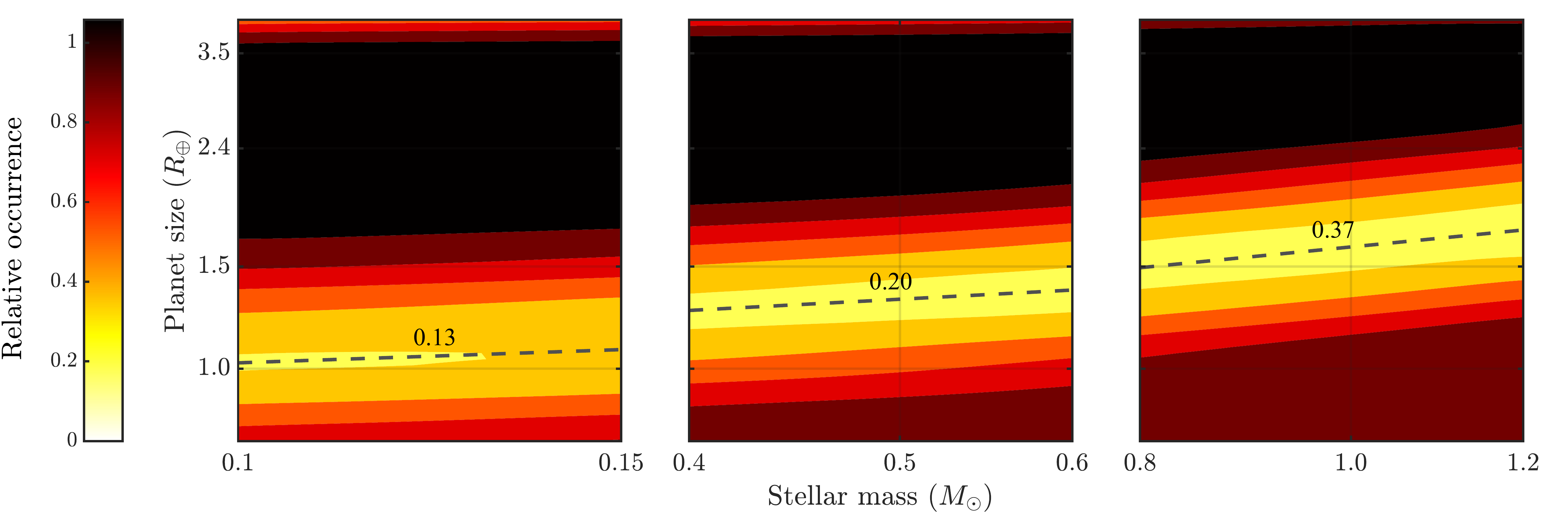}
\caption{Same as \Cref{fig:Rp_vs_Ms_distribution} but presenting an expanded view of the radius valley in three narrower stellar mass bins that span (from left to right) 0.1 to 0.15 $M_\odot$, 0.4 to 0.6 $M_\odot$ and 0.8 to 1.2 $M_\odot$. This plot clearly shows how the radius valley slope decreases with decreasing stellar mass. Dashed lines are based on our analytical estimate for the slope of the radius valley and show an excellent match with the simulation results. These slopes $\text{d log}R_p/\text{d log}M_\ast =$ 0.13, 0.20 and 0.37 for the three stellar mass bins as depicted in the figure.}
\label{fig:Rp_vs_Ms_distribution_different_ranges}
\end{figure*}

\Cref{fig:luminosity_and_mass_effect} also demonstrates how the location of the radius valley changes with stellar mass. When comparing the bottom panels for the two stellar masses, we can see that the radius valley around the less massive star moves to smaller planet sizes. This shift in the location of the radius valley is even more apparent in \Cref{fig:Rp_vs_Ms_distribution} where we show the planet size distribution as a function of planet size and stellar mass. This plot shows not just that the radius valley shifts to lower planet sizes with decreasing stellar mass, i.e., the slope of the radius valley is positive in the planet size-stellar mass space, but predicts that towards lower mass stars the slope of the radius valley becomes shallower. In \Cref{fig:Rp_vs_Ms_distribution_different_ranges}, we focus on three stellar mass bins that have an equivalent span in logarithmic space: 0.1 - 0.15 $M_\odot$, 0.4 - 0.6 $M_\odot$, and 0.8 - 1.2 $M_\odot$ (from left to right).

The positive slope of the radius valley in the $R_p - M_\ast$ space can be understood by same physics that we used to explain the radius valley slope in the $R_p-P$ space. As shown earlier, there is a positive correlation between stellar mass and planetary equilibrium temperature/insolation flux for a given orbital period $T_{eq} \propto M_\ast^{(\zeta/4) - (1/6)}$ \citep[e.g.][]{salaris2005a,eker2018a,berger2020b}. Therefore, at a particular orbital period, a planet around a low mass star is at a lower $T_{eq}$ as it receives less bolometric insolation flux ($S_p \propto T_{eq}^{1/4}$) in comparison to a planet at the same orbital period around a high mass star. In other words, a planet at a particular orbital period around a more massive star is more susceptible to a complete atmospheric loss. This explains why the radius valley moves to larger planet sizes around more massive stars, resulting in a positive slope in the $R_p-M_\ast$ space. 

We can obtain an analytical estimate for $\text{d log}R_p/\text{d log}M_\ast$ using \Cref{eq:slope_Rp_misc} by assuming that the orbital period distribution is independent of stellar mass \citep[e.g.][]{fulton2018a,petigura2022a} and that the short-period planets typically have similar core compositions, such that $\rho_*$ is independent of $P$ and $M_*$, \citep[e.g.][]{gupta2019a,doyle2019a,bower2019a,rogers2021a}, which yields
\begin{equation}\label{eq:slope_Rp_Ms}
    \frac{\text{d log}R_p}{\text{d log}M_\ast} = \frac{3 \zeta - 2}{36}.
\end{equation}
$\zeta$ has been observed to be $\gtrsim 2$ for stars more massive than 0.179 $M_\odot$ \citep[e.g.][]{salaris2005a,eker2018a}. \Cref{eq:slope_Rp_Ms} thus shows that for typical values of $\zeta$, the radius value has a positive slope. Furthermore, this equation encapsulates how the slope of the radius valley in the $R_p-M_\ast$ space depends on the correlation between stellar mass and luminosity ($L_\ast/L_{\sun}=(M_\ast/M_{\sun})^{\zeta}$.). For a larger (smaller) $\zeta$, the slope is larger (smaller).

It is commonly assumed that for zero-age main-sequence stars $\zeta \sim 3.5$, as can be shown using order-of-magnitude arguments. However, this is not accurate for actual stars - especially those in the mass range of 0.1 to 1.5 $M_\odot$. Even though this is a good approximation as it is within a factor of 2-3 of observed values, a change in $\zeta$ from 2 to 6 is equivalent to changing the radius valley slope estimate in the $R_p - M_\ast$ space from $\text{d log}R_p/\text{d log}M_\ast =$ 0.11 to 0.44. This is a significant change. In fact, $\zeta$ does vary in the aforementioned range as has been noted in past studies \citep[e.g.][]{salaris2005a,berger2020b}. For instance, in \citet{eker2018a}, the authors investigated the stellar mass-luminosity relation for a sample of 509 main-sequence stars within the mass range 0.179 $\leq M_\ast / M_\odot \leq $ 31. For stars with masses in the range 0.179 $\leq M_\ast/M_\odot \leq$ 0.45, 0.45 $<M_\ast / M_\odot \leq$ 0.72, and 0.72 $<M_\ast / M_\odot \leq$ 1.05, they found that $\zeta \simeq$ 2.0, 4.6 and 5.7, respectively.

\begin{table}
\centering
{
\renewcommand{\arraystretch}{1}
\begin{tabular}{c  c  c} 
\hline
Stellar mass bins ($M_\odot$) & $\zeta$\;($L_\ast \propto M_\ast^\zeta$) & $\text{d log}R_p/\text{d log}M_\ast$ \\ 
\hline

0.1 - 0.15 & 2.2 & 0.13\\

0.4 - 0.6 & 3.1 & 0.20\\

0.8 - 1.2 & 5.1 & 0.37\\

\hline

\end{tabular}
\caption{Summary of how the slope of the radius valley in planet size-stellar mass space, $\text{d log}R_p/\text{d log}M_\ast$, (last column) and the stellar mass-luminosity relation, $\zeta$, (middle column) change with stellar mass, for three stellar mass bins (first column).
}
\label{table:zeta_beta}
}
\end{table}

To track the luminosity evolution of stars included in \Cref{fig:Rp_vs_Ms_distribution_different_ranges}, we used MIST tracks. We find that $\zeta$ values for 3 Gyr old stars in the mass ranges shown in \Cref{fig:Rp_vs_Ms_distribution_different_ranges}, i.e., 0.1-0.15 $M_\odot$, 0.4-0.6 $M_\odot$ and 0.8-1.2 $M_\odot$ are $\sim$ 2.2, 3.1 and 5.1, respectively. Unsurprisingly, these values encompass the same range as found in \citet{eker2018a}. This is expected as the MIST datasets are derived from the stellar evolution code MESA, which is, in turn, tuned to match the properties of the observed stars \citep[e.g.][]{paxton2011a,paxton2013a}. Nevertheless, the fact that $\zeta$ increases with stellar mass in the mass range 0.1-1.2 $M_\odot$ implies that the slope of the radius valley in the planet size-stellar mass space increases in magnitude with increasing stellar mass. Consequently, given \Cref{eq:slope_Rp_Ms}, we find that the slope of the radius valley $\text{d log}R_p/\text{d log}M_\ast =$ 0.13, 0.20 and 0.37 for 3 Gyr old stars with masses 0.1-0.15 $M_\odot$, 0.4-0.6 $M_\odot$ and 0.8-1.2 $M_\odot$, respectively. We summarize this result in \Cref{table:zeta_beta}. These slopes have been plotted in \Cref{fig:Rp_vs_Ms_distribution_different_ranges} using dashed lines and provide an excellent match to the simulations.

\subsection{Comparing the predictions of core-powered mass-loss with photoevaporation, and observations}

\begin{figure}
\centering
\includegraphics[width=0.49\textwidth,trim= 210 340 900 290,clip]{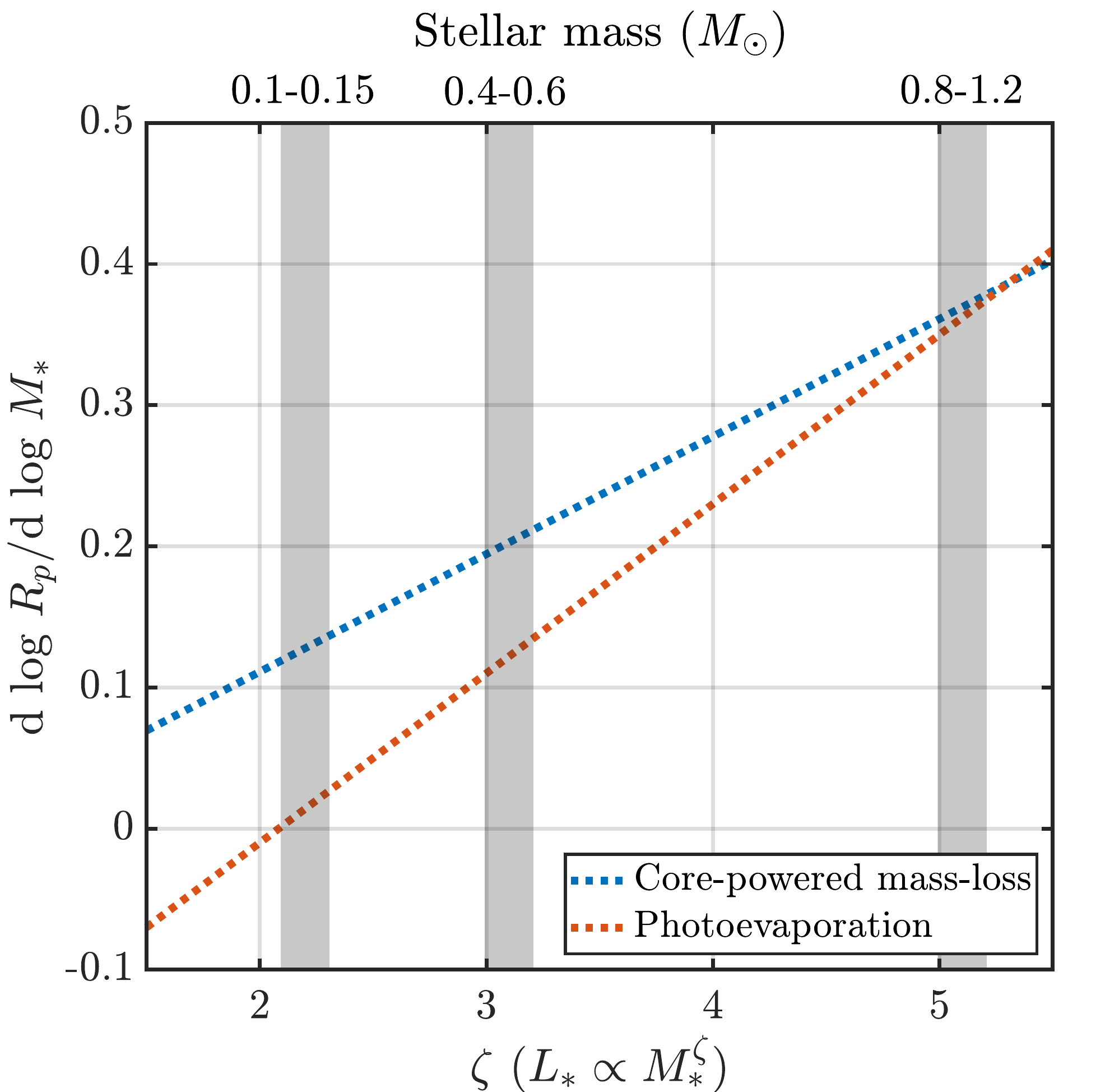}
\caption{Comparison between the predictions of core-powered mass-loss and photoevaporation models for the variation in the radius valley slope in planet size-stellar mass space ($L_\ast \propto M_\ast^\zeta$) as a function of $\zeta$ (bottom x-axis) and stellar mass (top x-axis). This figure demonstrates that the two mass-loss mechanisms predict increasingly different values for the slope of the radius valley in planet size-stellar mass space. Therefore, characterising the slope of the radius valley as a function of stellar mass around low mass stars may provide a fruitful avenue for distinguishing between core-powered mass loss and photoevaporation models.
Two observational studies, \citet[][]{berger2020b} and \citet[][]{petigura2022a}, have measured the value of this slope as $0.26^{+0.21}_{-0.16}$ and $0.18^{+0.08}_{-0.07}$, respectively. Both these estimates however can't be directly compared with the predictions here as (1) $\zeta$ for the relevant (survey) stellar mass range are not known and (2) these values correspond to the slope of the valley for the entire range of $\sim$ 0.5-1.5 $M_\odot$ and not for smaller stellar mass bins. Nevertheless, these observed slopes are in the range expected from evolution under core-powered mass-loss $\sim 0.20-0.35$. In future studies, observers could compare how the radius valley slope in the planet size-stellar mass space changes with $\zeta$ which could then give us further proof if atmospheric escape dictates planet evolution, and if so indeed, which among photoevaporation or core-powered mass-loss, if any, is the primary driver of this process.}
\label{fig:Rp_vs_Ms__slope_vs_zeta}
\end{figure}

The results presented above raise the question if these new predictions for the core-powered mass-loss mechanism are significantly different from those expected from photoevaporation. \Cref{fig:Rp_vs_Ms__slope_vs_zeta} aims to address this question. In this figure, we examine how the radius valley slope in the planet size-stellar mass space, $\text{d log}R_p/\text{d log}M_\ast$, changes with $\zeta$ (i.e. stellar mass) for the two mass-loss models. To estimate the slope of $\text{d log}R_p/\text{d log}M_\ast$ for the core-powered mass-loss theory, we used \Cref{eq:slope_Rp_Ms}; for estimating the same slope for photoevaporation, we used derivations from \citet{rogers2021a} where it was shown that $\text{d log}R_p/\text{d log}M_\ast \sim 0.12(\zeta -(2/3)) - 0.17$. For a full derivation of the latter result, the reader is referred to the discussion accompanying Equations 6 and 24 in \citet[][]{rogers2021a}. We find that for photoevaporation too the slope of the radius valley is likely to decrease in magnitude towards lower stellar masses because of the dependence of $\zeta$ on stellar mass. However, this figure also shows that while the difference in the slopes is quite small for solar-mass stars, it increases substantially towards lower stellar masses. This result further motivates efforts toward characterizing the radius valley around M dwarfs. 

Two observational studies, \citet[][]{berger2020b} and \citet[][]{petigura2022a}, have provided constraints on the radius valley slope in the planet size-stellar mass space: $0.26^{+0.21}_{-0.16}$ and $0.18^{+0.08}_{-0.07}$, respectively.
In theory, one could compare the predictions presented in \Cref{fig:Rp_vs_Ms__slope_vs_zeta} with these values. However, $\zeta$ for these studies were not estimated and it is not known how these slopes change with stellar mass. Although this implies that a direct comparison of the slopes estimated by \citet{berger2020b} and \citet{petigura2022a} and our predictions is not very informative, the magnitude of the observed slopes are in the range expected for planets evolving under core-powered mass-loss $\sim 0.20-0.35$. It should be noted that for correctly comparing a slope obtained from observational studies to the predictions of core-powered mass-loss or photoevaporation, one cannot simply choose a value for $\zeta$ using the MIST tracks. The purpose of using MIST tracks in this study is to simply demonstrate how the radius valley slope changes with stellar mass by using an open-source tool widely used in the community. The MIST tracks, on their part, are informed by theory and observations, however, the samples of stars used to constrain MESA is likely going to be different from the survey used for analyzing the planet demographics. For instance, to estimate $\zeta$ in this study we used an underlying host star distribution that is log-uniform in mass, has a fixed metallicity ($Z=0.014$) and age (3 Gyr), and identical planet multiplicity. These parameters may not be representative for a typical survey and thus the exact planet-host star distribution will impact the value of $\zeta$. Furthermore, the value of $\zeta$ should be different for volume-limited and magnitude-limited surveys. In the latter case, observations are biased towards stars that are more luminous for a certain mass. Therefore, $\zeta$ for a magnitude-limited survey is likely to be higher than the $\zeta$ for a volume-limited survey. In a nutshell, to compare the predictions of different theories with observations, future observational studies should estimate $\zeta$ using their actual stellar sample.

\section{Conclusions} \label{sec:Conclusions}


In this work, we explore how the core-powered mass-loss mechanism shapes the radius valley for planets orbiting low mass stars such as M-dwarfs. For this purpose, we extended our previous work where we had investigated the influence of this mechanism on planets hosted by FGK dwarfs \citep[][]{ginzburg2018a,gupta2019a,gupta2020a,gupta2021a} by accounting for the luminosity evolution of lower mass stars, which can be especially significant for stars such as TRAPPIST-1. We simulated populations of millions of planets with an orbital period distribution based on observations \citep[e.g.][]{petigura2022a} but a planet mass distribution that is log-uniform. Choosing a log-uniform planet mass distribution allowed us to be agnostic to the yet unknown `true' planet mass distribution given the uncertainties in how it scales with stellar mass.

We find that accounting for the evolution in the luminosity of FGK dwarfs instead of assuming a constant luminosity results in a negligible change in the size distribution of planets. This demonstrates that our assumption of a constant luminosity to model core-powered mass-loss for planets residing around FGK dwarfs in previous studies \citep[e.g.][]{ginzburg2018a,gupta2020a} was well justified. For stars of much lower masses, e.g. for a 0.1 $M_\odot$ star, we find however that the radius valley is narrower and less empty when we correctly account for stellar luminosity evolution. Nevertheless, the changes in the planet distribution are not significant when compared with the case where the luminosity is assumed to be constant in time. For instance, both models still predict the radius valley to be centered at roughly the same location in planet size and find the same slopes for the radius valley in planet-size orbital-period space.

In the planet size-orbital period space, we predict that the radius valley should have a negative slope of $\text{d log}R_p/\text{d log}P \simeq -0.11$ even around stars as small as 0.1 $M_\odot$, just as previously observed for FGK dwarfs \citep[e.g.,][]{vaneylen2018a,martinez2019a} and more recently, even for M-dwarfs \citep[e.g.,][]{vaneylen2021a,petigura2022a}. The valley is however moves to smaller planet size around a lower mass star and in addition, is narrower and less empty. Furthermore, our results show that the slope of the radius valley in the planet size-stellar mass parameter space does not stay constant with stellar mass but decreases in magnitude towards lower mass stars. Previously, we predicted a slope of $\text{d log}R_p/\text{d log}M_\ast \sim$ 0.34 around Sun-like stars, i.e. stars close to 1.0 $M_\odot$; see \citet{gupta2020a}. Because of the core-powered mass-loss' dependence of stellar luminosity, we demonstrate analytically and numerically that, if core-powered mass-loss is indeed dictating planet evolution, the radius valley slope should decrease to $\text{d log}R_p/\text{d log}M_\ast \sim$ 0.20 around early M-dwarfs, i.e. stars close to 0.5 $M_\odot$, and to $\sim$0.13 around late M-dwarfs, i.e. those with masses close to 0.1 $M_\odot$. As shown in this work, the slope of the radius valley in the planet size-stellar mass space is $\text{d log}R_p/\text{d log}M_\ast = (3 \zeta - 2)/36 $, where $L_\ast/L_{\sun}=(M_\ast/M_{\sun})^{\zeta}$. For a larger (smaller) value of $\zeta$, the slope in planet size-stellar mass space is also larger (smaller). Based on MESA-derived stellar evolution tracks, we find that $\zeta$ increases with stellar mass in the range $\sim$ 0.08-1.5 $M_\odot$. We thus find that the slope of the radius valley also increases (decreases) with increasing (decreasing) stellar mass and that the magnitude of the increase (decrease) can be directly calculated from from the mass-luminosity relation ($L_\ast/L_{\sun}=(M_\ast/M_{\sun})^{\zeta}$.). Finally, we demonstrated that core-powered mass-loss and photoevaporation predict increasingly different slopes for the radius valley in the planet size-stellar mass space as one moves towards lower mass stars. Taken together, these findings provide us with new avenues to distinguish the signatures of core-powered mass-loss and photoevaporation observationally.

In this work, we have explored how the radius valley evolves around low mass stars if atmospheric evolution of planets is solely dictated by the core-powered mass-loss mechanism. Our results and predictions motivate the need for new observational studies and surveys that search for planet populations around low mass stars. Studies such as \citet[][]{cloutier2020a}, \citet[][]{vaneylen2021a} and \citet[][]{petigura2022a} show some of the recent progress that has been made in this direction. Extending such studies and surveys further to even lower stellar masses and larger datasets with the help of TESS \citep[e.g.][]{tess2015a}, CHEOPS \citep[e.g.][]{cheops2021a}, PLATO \citep[e.g.][]{plato2014a} and other ground and space-based instruments, could accelerate efforts towards understanding which mechanism, if any, is the true cause shaping the radius valley and dictating the evolution of small exoplanets.

\section*{Acknowledgements}

{We thank the anonymous referee for valuable comments that helped improve the manuscript.} A.G. is supported by the Future Investigators in NASA Earth and Space Science and Technology (FINESST) grant 80NSSC20K1372. L.N. acknowledges support from the University of California - Leadership Excellence Through Advanced Degrees (UC LEADS). H.E.S. gratefully acknowledges support from NASA under grant number 80NSSC21K0392 issued through the Exoplanet Research Program.


\section*{Data Availability}
The data underlying this article will be shared on reasonable request to the corresponding author.


\bibliographystyle{mnras}
\bibliography{planet_evo}








\bsp	
\label{lastpage}
\end{document}